\documentclass{iopart}
\usepackage{amssymb}
\usepackage{iopams}
\usepackage{amsfonts}
\catcode`\@=11
\renewcommand\footnoterule{%
  \kern-3\p@
  \hrule\@width.4\columnwidth
  \kern2.6\p@}
\renewcommand\@makefntext[1]{%
    \parindent 1em\noindent
    \hb@xt@1.8em{\hss$^{\@thefnmark}$)}\hspace{2pt}%
    \footnotesize\rmfamily#1}  
\def\@makefnmark{\hspace{.5pt}\hbox{$^{\@thefnmark}$%
\hspace{-1pt})}} 
\catcode`\@=11
\renewcommand\footnoterule{%
  \kern-3\p@
  \hrule\@width.4\columnwidth
  \kern2.6\p@}
\renewcommand\@makefntext[1]{%
    \parindent 1em\noindent
    \hb@xt@1.8em{\hss$^{\@thefnmark}$)}\hspace{2pt}%
    \footnotesize\rmfamily#1}  
\def\@makefnmark{\hspace{.5pt}\hbox{$^{\@thefnmark}$%
\hspace{-1pt}}} 

\newcommand{\dd}{\dagger}

\def\RR{\mathbb{R}}

\def\CC{\mathbb{C}}

\def\p{\partial}
\def\a{\alpha}

\def\d{\delta}

\def\G{\Gamma}
\def\sg{\sigma}

\def\om{\omega}

\newcommand{\cC}{\mathcal{C}}
\newcommand{\cD}{\mathcal{D}}

\newcommand{\cH}{\mathcal{H}}

\newcommand{\cK}{\mathcal{K}}
\newcommand{\cM}{\mathcal{M}}
\newcommand{\cN}{\mathcal{N}}
\newcommand{\cP}{\mathcal{P}}
\newcommand{\cT}{\mathcal{T}}
\newcommand{\cQ}{\mathcal{Q}}
\newcommand{\cR}{\mathcal{R}}

\newcommand{\bfomega}{\boldsymbol{\omega}}

\newcommand{\bP}{\mathbf{P}}

\newcommand{\bT}{\mathbf{T}}

\newcommand{\fk}{\mathfrak{k}}
\newcommand{\ffl}{\mathfrak{l}}
\newcommand{\fp}{\mathfrak{p}}
\newcommand{\fq}{\mathfrak{q}}

\newcommand{\be}[1]{\begin{equation}\label{#1}}
\newcommand{\ee}{\end{equation}}
\newcommand{\ba}[1]{\begin{eqnarray}\label{#1}}
\newcommand{\ea}{\end{eqnarray}}
\newcommand{\rf}[1]{(\ref{#1})}
\newcommand{\nn}{\nonumber}

\newcommand{\sign}{\mbox{\rm sign}\,}
\newcommand{\diag}{\mbox{\rm diag}\,}
\renewcommand{\span}{\mbox{\rm span}\,}
\newcommand{\ra}{\rangle}
\newcommand{\la}{\langle}

\begin{document}
\title[$\cP\cT-$symmetry, Cartan decompositions, Lie triple systems and Clifford algebras]%
{$\cP\cT-$symmetry,  Cartan decompositions, Lie triple systems and Krein space related Clifford  algebras}

\author{U. G\"{u}nther${}^a$\  and S. Kuzhel${}^b$}

\address{${}^a$\ Research Center Dresden-Rossendorf, POB 510119, D-01314
Dresden, Germany\\
${}^b$\ Institute of Mathematics of the NAS of Ukraine,
01601 Kyiv, Ukraine}
\eads{\mailto{u.guenther@fzd.de},\ \mailto{kuzhel@imath.kiev.ua}}

\begin{abstract}
Gauged $\cP\cT$ quantum mechanics (PTQM) and corresponding Krein
space setups are studied. For models with constant non-Abelian gauge potentials and extended parity inversions compact and noncompact Lie group components are analyzed via Cartan decompositions. A Lie-triple structure is found and an interpretation as $\cP\cT-$symmetrically generalized Jaynes-Cummings model is possible with close relation to recently studied cavity QED setups with transmon states in multilevel artificial atoms. For models with Abelian gauge potentials a hidden Clifford algebra structure is found and used to obtain the fundamental symmetry of Krein space related $J-$selfadjoint extensions for PTQM setups with ultra-localized potentials.
\end{abstract}
\pacs{02.20.Sv, 02.30.Tb, 03.65.Ca, 03.65.Db, 42.50.Pq, 11.30.-j}
\ams{47B50, 46C20, 81Q12, 15A66, 20N10, 17B81}\mbox{}\\[1ex]

During the last ten years many of the basic features of
quantum mechanics with $\cP\cT-$symmetric Hamiltonians (PTQM)
\cite{B1,B4} have been worked out in detail and are now to a certain
degree well understood. This concerns the mapping of the
PTQM-sector of exact $\cP\cT-$symmetry (The Hamiltonian $H$ and
its eigen(wave)functions are both $\cP\cT-$symmetric and the
spectrum of $H$ is purely real.) to conventional (von-Neumann)
quantum mechanics with Hermitian Hamiltonians \cite{M2}, the
relevance of the $\cC-$operator as dynamically adapted mapping
\cite{B-C} between Krein-space-related indefinite metric
structures \cite{AGK} and positive definite metrics of usual
Hilbert spaces (required for a sensible probabilistic
interpretation of the related wave functions) as well as the
understanding of $\cP\cT-$symmetric Hamiltonians as selfadjoint
operators in Krein-spaces \cite{Ja,AK2,LT,GSZ,TT}.

Here, we will discuss some up to now unnoticed structural links of
PTQM, and Krein space related models in general, to Lie algebra
and Lie group related Cartan decompositions
\cite{knapp,gilmore-old,gilmore-new}, Lie triple systems
\cite{lister-LTS,harris-LTS,bertram-LNM,hodge-LTS,bertram-didry,izq-jpa2010}
as well as to Clifford algebras \cite{cliff-book}. Identifying
these underlying structures will help in recognizing hidden
$\cP\cT-$like involutory structures in physical models which are
up to now not related with $\cP\cT-$symmetry and to deeper
understand these models and the role of $\cP\cT-$symmetry in
general.

We start from the simplest $\cP\cT-$symmetric Hamiltonian $H$,
$[\cP\cT,H]=0$, of differential operator type
\ba{2}\fl
H=p^2+V(x)&,\quad & p:=-i\p_x,\quad V(-x)=V^*(x),\quad\cP
x\cP=-x,\quad \cP p\cP=-p\nn\\
&&\cT iI\cT=-iI,\quad \cT x\cT=x,\quad \cT p\cT=-p.
\ea
In general, this Hamiltonian is a $\cP-$selfadjoint operator in a
Krein-space $(\cK,[\cdot,\cdot]_\cP)$ (see, e.g., \cite{AZ,DL})
with $[\cdot,\cdot]_\cP:=(\cdot,\cP\cdot)$ being the $\cP\cT$
inner product \cite{B4}, $[H\phi,\psi]_\cP=[\phi,H\psi]_\cP$, i.e.
\be{2a}
\cP H=H^\dd \cP.
\ee
Because of $\cP p=-p\cP= -p^\dd\cP$, i.e.
$[p\phi,\psi]_\cP=-[\phi,p\psi]_\cP$, this $\cP-$selfadjointness
is spoilt for the gauged Hamiltonian\footnote{The same holds for
any $H=p^2+a(x)p+\tilde V(x)$ with $\tilde V(x)=\tilde V^*(-x)$
and sufficiently smooth $a(x)=a^*(-x)$ --- as rewritten version of
$H_g$. Compared to standard physics notations $H_g=(p-eA)^2+V(x)$
we absorbed the charge $e$ into the gauge potential $eA\mapsto
A$.}
\ba{2g} H_g&=&(p-A)^2+V(x), \qquad A(-x)=A^*(x)\label{2g}\\
\cP H_g&\neq &H_g^\dd \cP.\label{2g-b}
\ea
Instead the gauge transformation (Kummer-Liouville transformation
\cite{polyanin-zaitsev})
\be{3}
U:\ H_g\mapsto H=UH_gU^{-1}
\ee
\be{3b}
U=e^{-i\int_0^xA(s)ds}
\ee
together with \rf{2a}, $\cP=\cP^\dd$ and
$\left[U^\dd\right]^{-1}=\left[U^{-1}\right]^\dd$ leads to the
pseudo-Hermiticity condition
\be{4}
\eta H_g=H_g^\dd \eta, \qquad \eta:= U^\dd \cP U,\quad
\eta=\eta^\dd.
\ee
$\cP\cT-$symmetry of the system remains preserved under the gauge
transformation $U$
\be{6-0}
[\cP\cT, U]=0, \qquad  [\cP\cT,H_g]=0, \quad [\cP\cT,H]=0.
\ee
These facts are well known and have been widely discussed for
various PTQM models \cite{AH,MO,BQ,KS1,fring-gauge,JO}.

Next we assume, for simplicity, a purely real coordinate
dependence $x\in \Omega\subseteq \RR$ with $\Omega$ any
$\cP-$symmetric interval. Then splitting $A(x)=A_+(x)+iA_-(x)$
into even and odd components, $\cP A_\pm(x)=A_\pm(-x)=\pm
A_\pm(x)$, leads to a factorization of $U$ into unitary and
Hermitian $\cP-$selfadjoint factors
\ba{5}
&&U=U_uU_h,\qquad U_u=e^{-i\int_0^xA_+(s)ds},\qquad
U_h=e^{\int_0^xA_-(s)ds}\label{5a}\\
&&U_u^\dd=U_u^{-1},\qquad U_h^\dd=U_h\nn\\
&&\cP U=U^\dd \cP,\qquad \cP U_u=U_u^\dd \cP,\qquad \cP
U_h=U_h\cP.\label{5b}
\ea
This is just the simplest (Abelian) version of a polar
decomposition which here is naturally associated with the
corresponding decomposition of the metric $\eta=J|\eta|$ into
modulus $|\eta|:=\sqrt{ \eta^2}=U_h^2$ and involution
$J:=\eta|\eta|^{-1}=U_u^{-1}\cP U_u=J^\dd=J^{-1}$. It shows that
$H_g$ is $J-$selfadjoint in the weighted ($|\eta|-$deformed)
Hilbert space $L_2(|\eta|dx)$ with inner product
$(\phi,\psi)_{|\eta|}: =\int_\RR
\psi(x)\phi^*(x)e^{2\int_0^xA_-(s)ds}dx$
\be{6}
(H_g\phi,J\psi)_{|\eta|}=(\phi,JH_g\psi)_{|\eta|}\,.
\ee
Obviously, the unitary component $U_u$ of the gauge transformation
$U(x)$ rotates the original involution (Krein space metric) $\cP$
into the new involution $J=U_u^{-1}\cP U_u$ whereas the Hermitian
component $U_h$ induces the new integration weight $|\eta|$, i.e.
we have a Krein space mapping $U:\ (\cK_\cP,[.,.]_\cP)\mapsto
(\tilde \cK_J,[.,.]_{|\eta|J})$.

A further mapping $\rho$ will be needed to pass from
$L_2(|\eta|dx)$ in \rf{6} to a Hilbert space $\cH$ where a
Hamiltonian $H_g$ with real spectrum (exact $\cP\cT-$symmetry)
will be not only $J-$selfadjoint but selfadjoint \cite{M2,SH}. The
intensive studies during the last years showed that this $\rho$
will strongly depend on the concrete form of the
$\cP\cT-$symmetric potentials $A(x)=A^*(-x)$, $V(x)=V^*(-x)$ and,
in general, it will be highly nonlocal \cite{B4,M-metric}.
Subsequently, we mainly concentrate on the symmetry structures
inherent in the model and we will not focus on the
nonlocalities\footnote{The nonlocalities are of similar type as
those arising in a Foldy-Wouthuysen transformation when a Dirac
Hamiltonian is block-diagonalized \cite{bjorken,thaller}.} as the
latter are typical, e.g., for the construction of $\cC$ operators
for Hamiltonians built over differential operators \cite{BK}.

The above decomposition \rf{5a} indicates on two ways of possible
model generalizations based (i) on a generalization of the Abelian
gauge potential to a non-Abelian one or, via slightly different
structures, on (ii)
the direct use of a hidden Clifford algebra.\\

\noindent {\bf Non-Abelian gauge potentials, Cartan decompositions
and Lie triple systems} First we note that the decomposition
\rf{5a} of the gauge transformation $U$ into unitary and Hermitian
components can be regarded as trivial Abelian version of a Cartan
decomposition of a Lie group into a compact subgroup and a
noncompact homogeneous coset space. Subsequently we demonstrate
the interrelation of $\cP\cT-$symmetry and  Cartan decompositions
of Lie groups (and Lie algebras) on the simplest example  of a
matrix Hamiltonian with non-Abelian but constant\footnote{In case
of non-Abelian local (coordinate dependent) gauge potentials in
theories over a space-time manifold $\cM$ (e.g. over usual
Minkowski space) finite gauge transformation operators $U$ will
have the form of path-ordered exponentials
\cite{wilson-line1,wilson-line2}. For simplicity we restrict our
consideration here to constant gauge transformations only.
Regardless of the gauge field type (constant or coordinate
dependent) in case of $x\in \Omega\subseteq \RR$ the spectral
problem can be regarded as matrix-Sturm-Liouville problem
\cite{sturm-liouv-book}.} gauge potential $A$. The parity
inversion $\bP$ we assume of tensor product type, i.e. we set for
our model
\ba{6-1}
\fl &&H_g=(p-A)^2+V(x),\quad A\in \CC^{m\times m}, \quad V(x)\in
\CC^{m\times m}\otimes L_1(\RR)\label{6-1-1}\\
\fl &&\left[\bP\cT,H_g\right]=0, \quad \bP=\Theta \otimes \cP,
\quad \Theta\in\RR^{m\times m},\quad \Theta^2=I_m, \qquad
\bP^2=I_m\otimes I\label{6-1-2}.
\ea
Involution property $\Theta^2=I_m$ and reality
$\Theta\in\RR^{m\times m}$ imply diagonalizability and symmetry of
the matrix $\Theta=\Theta^T$. This means that without loss of
generality, i.e. modulo a global $SO(m,\RR)$ rotation, we may fix
henceforth $\Theta=I_{p,q}=\diag(I_p,-I_q)$, $p+q=m$. Furthermore,
we assume for simplicity that $\cT$ acts as the same complex
conjugation as for the scalar Hamiltonian \rf{2g}, i.e. $\cT\cong
I_m\otimes\cT$ so that involution commutativity concerning the
extended parity inversion $\bP$ is fulfilled trivially\footnote{In
general, the time involution $\cT$ may be extended nontrivially to
any anti-linear involution $\bT=\mu\otimes \cT$ with $\mu^2=I_m$,
$\mu\in\CC^{m\times m}$. In the simplest case  of
$\mu\in\RR^{m\times m}$, involution commutativity $[\bP,\bT]=0$
together with fixed $\Theta=I_{p,q}$ implies a block-diagonal
$\mu=\diag(\mu_p,\mu_q)=S I_{r,s}S^{-1}$, $S\in SO(m,\RR)$ with a
possibly different signature $(r,s)\neq (p,q)$. Moreover, even
involution commutativity may be violated, $[\bP,\bT]\neq 0$ as,
e.g., for the pinor-representations \cite{cpt-review} of the Dirac
equation. We leave corresponding considerations to future research
and restrict our attention here to the simplest ansatz
$\bT=I_m\otimes\cT$ only.} \ $[\bP,\cT]=0$. In this case
$\bP\cT-$symmetry, $[\bP\cT,H_g]=0$, implies
\ba{6-2}
\Theta A^* \Theta=A,\qquad \Theta V^*(-x)\Theta=V(x)
\ea
whereas $\bP-$selfadjointness  $\bP H^\dd\bP=H$ of the globally
re-gauged Hamiltonian
\be{6-3}
H=UH_gU^{-1}=p^2+e^{-iAx}V(x)e^{iAx},\qquad U=e^{-iAx}
\ee
leads to the additional conditions
\ba{6-4}
\Theta A^\dd \Theta=-A,\qquad \Theta V^\dd(-x)\Theta=V(x).
\ea
Together \rf{6-2} and \rf{6-4} give
\be{6-6}
A=-A^T, \qquad V=V^T,
\ee
and they fix via \rf{6-3} the Lie group structure of the gauge
transformation $U$. Denote the set of corresponding Lie group
elements by $G_\Theta\ni U$ and the vector space of its Lie algebra elements
by $g_\Theta$. Then for the elements $a\in g_\Theta$, because of
$a:=-iA$,  it holds
\be{6-5}
a=-a^T, \qquad \Theta a^\dd \Theta=a.
\ee
Hence, $g_\Theta$ is constituted by the $\Theta-$Hermitian
elements of $so(m,\CC)$. In order to understand the role of this
$\Theta-$Hermiticity condition we first note that the compact
subgroup of the special complex orthogonal group $SO(m,\CC)$ is
the real orthogonal group $SO(m,\RR)$, whereas the (homogeneous)
coset space $SO(m,\CC)/SO(m,\RR)$ parameterizes the noncompact
("boost"-type) transformations. This is well known (see, e.g.
\cite{gilmore-old}, chapt. 9, sect. II) and follows trivially from
the Cartan de\-com\-po\-si\-tion of general $GL(m,\CC)$ matrices
into unitary compact components and Hermitian noncompact
components (i.e. from their polar decomposition). In fact, the
corresponding Cartan involution $\tau$ for the Lie algebra
$gl(m,\CC)\ni a$ is $\tau(a)=-a^\dd$ and $gl(m,\CC)$ can be
decomposed as $gl(m,\CC)=\fk\oplus\fp$ with $\tau \fk=\fk$, $\tau
\fp=-\fp$ for compact subalgebra $\fk$ and the set of noncompact
coset elements $\fp$, respectively. Imposing the additional
antisymmetry restriction $a=-a^T$ for $so(m,\CC)$ elements the
Cartan involution reduces to complex conjugation  $\tau
(a)=-a^\dd=a^*=\cT a$. Accordingly, $\cT$  splits $so(m,\CC)$ just
into real and purely imaginary components
\ba{6-6}
\fl&&so(m,\CC)=\fk\oplus\fp,\qquad \fk=so(m,\RR),\quad \fp=\{b\in
so(m,\CC)| b=if, f\in so(m,\RR)\}\label{6-6a}\\
\fl&& \cT \fk=\fk,\qquad \cT\fp=-\fp.\label{6-6b}
\ea
The $\Theta-$Hermiticity condition in \rf{6-5} refines this
decomposition by an additional $\Theta-$related block structure.
Explicitly $\Theta a^\dd\Theta=a$ implies
\ba{6-7}\fl &&a=:\left(
    \begin{array}{cc}
      iu & v \\
      -v^T & iw \\
    \end{array}
  \right),\qquad u\in \RR^{p\times p},\ v\in \RR^{p\times q},\ w\in \RR^{q\times
  q}\label{6-7-1}\\
\fl &&\fk_\Theta=\{b\in so(m,\RR)| b=\left(
           \begin{array}{cc}
             0 & v \\
             -v^T & 0 \\
           \end{array}
         \right)\},\label{6-7-2}\\
\fl && \fp_\Theta=\{c\in so(m,\CC)|c=if=\left(
                                                         \begin{array}{cc}
                                                           i u & 0 \\
                                                           0 & iw \\
                                                         \end{array}
                                                       \right),\ f\in so(p,\RR)\oplus
so(q,\RR)\}\label{6-7-3}\\
\fl && b^\dd=-b,\quad b\in\fk_\Theta, \qquad c^\dd=c,\quad
c\in\fp_\Theta \label{6-7-4}.
\ea
Denoting the Cartan decomposition of $su(p,q)$ by\footnote{Recall
that the compact subgroup of $SU(p,q)$ is $S(U(p)\times U(q))$ (see,
e.g. \cite{gilmore-old}).}
\be{6-7a}
su(p,q)=\ffl\oplus\fq,\qquad \ffl=s(u(p)\oplus u(q)),\qquad
\fq=su(p,q)\ominus \ffl
\ee
we see from $a=-iA$ with $A=-A^T$ and $\Theta A^\dd\Theta=-A$,
i.e. $A\in so(m,\CC)\cap su(p,q)$, that
\ba{6-7b}
&&g_\Theta=\{a\in so(m,\CC)| a=if,\ f\in  so(m,\CC)\cap su(p,q)\}=\fk_\Theta\oplus\fp_\Theta\nn\\
&& \fk_\Theta=so(m,\CC)\cap i\fq,\qquad \fp_\Theta=so(m,\CC)\cap
i\ffl.
\ea
This means that $g_\Theta$ can be considered as a "Wick rotated"
$so(m,\CC)\cap su(p,q)$, an $so(m,\CC)\cap su(p,q)$ with Weyl
unitary trick applied not only to the noncompact component $\fq$
but to the algebra as a whole. Correspondingly the roles of
compact and noncompact components in $su(p,q)\cap so(m,\CC)$ and
$g_\Theta$ are interchanged $\ffl,\fq \rightleftarrows
\fp_\Theta,\fk_\Theta$. The latter fact explains the
block-diagonal decomposition of the noncompact $\fp_\Theta$ in
\rf{6-7-2} and the off-diagonal block form of $\fk_\Theta$.

Next we note that the intersection set $g_\Theta$ is not a Lie
algebra itself. Rather this Lie algebra subspace $g_\Theta$ forms
a Lie triple system (LTS) (see, e.g. \cite{bertram-LNM}, sect. 1.1; \cite{izq-jpa2010}, sect. 10).
To see this we follow standard techniques
\cite{lister-LTS,harris-LTS,bertram-LNM,bertram-didry} and denote
by $\kappa$ the Lie algebra involution
\be{6-7bb}
\kappa(a):=-\Theta a^\dd\Theta.
\ee
Then the $\Theta-$Hermiticity condition in \rf{6-5} defines
$g_\Theta$ as $\kappa-$odd subspace in $so(m,\CC)$
\ba{6-7bc}
g_\Theta=\{a\in so(m,\CC)|\kappa(a)=-a\},
\ea
whereas the commutator $[g_\Theta,g_\Theta]$ is $\kappa-$even
$\kappa([g_\Theta,g_\Theta])=[g_\Theta,g_\Theta]$, i.e. $g_\Theta$
does not close under the Lie bracket
$[g_\Theta,g_\Theta]\nsubseteq g_\Theta$. It only closes under the
ternary composition\footnote{From the large number of recent
studies on ternary and $n-$ary Lie algebras as well as metric Lie
$3-$ and $n-$algebras we note as a few examples
\cite{izq-jpa2010,duplij,curtright,Figueroa-1}.}
\be{6-7bd}
a,b,c\in g_\Theta:\qquad [a,[b,c]]\in g_\Theta
\ee
so that $g_\Theta$ is indeed a Lie triple system (LTS)
$[[g_\Theta,g_\Theta],g_\Theta]\subseteq g_\Theta$.

For completeness, we display the Cartan decomposition of the group elements of
the set $G_\Theta= K_\Theta\Pi_\Theta $. Separately considered the
compact and the noncompact subset, $K_\Theta\subset SO(m,\RR)$ and
$\Pi_\Theta\subset SO(m,\CC)/SO(m,\RR)$, have parameterizations
induced by the corresponding Lie algebra elements in \rf{6-7-2},
\rf{6-7-3} (see e.g. \cite{gilmore-old}, chapt. 9, sect. IV)
\ba{6-8}
\fl K_\Theta&=&\{U_\fk\in SO(m,\RR)|\ U_\fk=e^{bx}=\left(
             \begin{array}{cc}
               \cos\left(\sqrt{vv^T}x\right) & v\frac{\sin\left(\sqrt{v^Tv}x\right)}{\sqrt{v^Tv}} \\
               -\frac{\sin\left(\sqrt{v^Tv}x\right)}{\sqrt{v^Tv}}v^T & \cos\left(\sqrt{v^Tv}x\right) \\
             \end{array}
           \right),\ b\in \fk_\Theta\},\nn\\ 
\fl\Pi_\Theta&=&\{U_\fp\in SO(m,\CC)/SO(m,\RR)|\ U_\fp=
e^{cx}=\diag(e^{iux},e^{iwx}),\ c\in \fp_\Theta\}.
\ea
Furthermore, it follows from \rf{6-7-4} that
\be{6-8-3}
U_\fk^\dd=U_\fk^{-1},\quad U_\fp^\dd=U_\fp
\ee
as generalization of decomposition \rf{5a} for the Abelian
gauge transformation.

In the trivial case of $\Theta=I_m$ there is no compact subgroup
present at all and the global gauge transformations $U$ are pure
boosts
\be{6-9}
U=e^{iux}=e^{-iAx}\in\Pi_I, \quad A=-A^T\in \RR^{m\times
m},\qquad U=U^\dd.
\ee
This fact is due to the obvious anti-Hermiticity of the gauge
potential $A=-A^\dd$ which is in clear contrast to the Hermitian
gauge potentials present in the Hermitian Hamiltonians of
conventional (von Neumann) quantum mechanics. For $m=2$, e.g., it
holds $iu=\a \sg_2x$, $\a\in\RR$ with $A=ia\sg_2$ so that
$U=e^{\a\sg_2 x}=\cosh(\a x)I_2+\sinh(\a x)\sg_2$ similar to
earlier findings e.g. in \cite{GS-pra,GS-prl}.

In contrast, for $\Theta\neq I_m$, $m\ge 2$ and vanishing
noncompact component, we find the gauge potentials $A$ as
antisymmetric Hermitian matrices $A\in i\fk_\Theta=\{A\in
so(m,\CC)|A=ib,\ b\in so(m,\RR)\}$. In the simplest case, $m=2$,
this reduces to $\Theta=\sg_3$, $A=\a \sg_2$, $\a\in\RR$ and
$U_\fk=e^{-i\a\sg_2}\in SO(2,\RR)\subset U(2)$.

For general $\Theta$ the gauge potential $A$ will be composed
simultaneously of anti-Hermitian as well as Hermitian components
corresponding to non-compact and compact components of the Lie
algebra element $a$, respectively.

The global gauge transformations $U\in G_\Theta$ are
$\bP\cT-$symmetry preserving
\be{6-10}
[\bP\cT,U]=0,\quad  [\bP\cT,H_g]=0,\quad [\bP\cT,H]=0,
\ee
in analogy to \rf{6-0} for Abelian systems. In contrast, the
$\bP-$symmetry properties of the $U\in G_\Theta$ components are
reversed compared to that for the Abelian $U$ in \rf{5b}
\be{6-11}
U\in G_\Theta:\qquad \bP U_\fk=U_\fk \bP,\qquad \bP
U_\fp=U_\fp^{-1}\bP.
\ee
This reversed behavior can be traced back to the special interplay
of complex conjugation and the antisymmetry of the gauge potential
as $so(m,\CC)$ element. On its turn it implies (via
$\bP-$Hermiticity of the re-gauged Hamiltonian $H$ in \rf{6-3},
the relation to the original Hamiltonian $H_g$, as well as
\rf{6-8-3}, \rf{6-11} and the decomposition $U=U_\fk U_\fp$) that
$H_g$ itself is $\bP-$Hermititian as well:
\ba{6-12}
\fl\bP H=H^\dd\bP\quad\Longrightarrow\quad \eta H_g=H_g^\dd
\eta,\qquad \eta=U^\dd \bP U=U_\fp U_\fk^{-1}\bP U_\fk U_\fp=\bP.
\ea
A simple explicit comparison of the $\cP-$ and
$\bP-$pseudo-Hermiticity conditions for the gauged Hamiltonians in
\rf{2g} and \rf{6-1-1} shows that the violation of the
$\cP-$Hermiticity for the scalar $H_g$ with Abelian gauge
potential is due to the non-vanishing
derivative term $i\p_x A(x)$ in $H_g$. The vanishing of this term
$i\p_x A=0$ for the constant (global) gauge potential $A$ removes
this obstruction and leads to preserved $\bP-$selfadjointness of
$H_g$ in \rf{6-1-1}, $[H_g\phi,\psi]_\bP=[\phi,H_g\psi]_\bP$.
Effectively, this results from the sign invariance of the
$Ap-$term under the simultaneous action of $\cP p=-p\cP$ and
$\Theta A=-A^\dd \Theta$ used for the construction of the Krein
space adjoint with regard to $[.,.]_\bP$\,.

Before we turn to the discussion of Clifford algebra related
structures in the $\cP\cT-$symmetric scalar Schr\"odinger
equation, we note that the $\bP\cT-$symmetric matrix Hamiltonian
$H_g$ in \rf{6-1-1} with constant gauge potential $A$ and
appropriately chosen $V(x)$ can be related to a Jaynes-Cummings
type Hamiltonian\footnote{For recent discussions of
Jaynes-Cummings models see, e.g.,
\cite{jaynes-cummings1,jaynes-cummings2}.} with additional
non-Hermitian $\bP\cT-$symmetric degrees of freedom. To see this
we introduce creation and annihilation operators
$d^\dd:=(-ip+x)/\sqrt2$, $d:=(ip+x)/\sqrt2$ and split the Lie
algebra element $a$ (see eq. \rf{6-7}) in strictly upper and lower
triangular (nilpotent) components
\be{t1}
a=c-c^T,\qquad c:=\left(%
\begin{array}{cc}
  i\tilde u & v \\
  0 & i\tilde w \\
\end{array}%
\right), \qquad c^m=0
\ee
with $\tilde u$, $\tilde w$ the strictly upper triangular components of $u$, $w$.
For
\be{t2}\fl
V(x)=(x^2-1)I_m+2(c+c^T)x +a^2+2\bfomega, \quad \bfomega=\diag [\om_1,\cdots,\om_m],\quad
\om_j\in\RR
\ee
and particle number operator $N=d^\dd d$ this yields, e.g.,
\be{t3}
\frac12 H_g=N+\sqrt2 (cd+c^Td^\dd)+\bfomega
\ee
describing a special type of $\bP\cT-$symmetry preserving
(gain-loss-balanced\footnote{For other $\cP\cT-$symmetric
gain-loss-balanced systems see, e.g.,
\cite{gain-loss1,gain-loss2,gain-loss3,gain-loss4,gain-loss5,gain-loss6}.})
$d-$particle-induced excitation process in a multi-level quantum
system. Models of this type can be considered, e.g., as
$\bP\cT-$symmetric generalization of the recently studied circuit
and cavity QED setups \cite{transmon1,transmon2} allowing for the
interaction of a single ($d-$)mode of the cavity electromagnetic
field with a set of transmon states
of a multilevel artificial atom with level energies $\om_j$\,.\\[1ex]

\noindent {\bf Krein space related hidden Clifford algebra} The
analysis of the scalar $\cP\cT-$symmetric Hamiltonian \rf{2g} with
local Abelian gauge potential $A(x)$ can be pursued in another
direction by concentrating on the symmetry properties of the
unitary factor $U_u=e^{-i\cQ}$, \ $\cQ:=\int_0^xA_+(s)ds$ in \rf{5}
which was responsible for the rotation of the involution as $U_u:\
\cP\mapsto J=U_u^{-1}\cP U_u$. Representing $\cQ$ as
\be{7}
\cQ=\cR q, \qquad \cR:=\sign(\cQ),\quad q:=|\cQ|
\ee
we see that the essential structure underlying the
$\cP-$Hermiticity condition $\cP U=U^\dd \cP$ together  with $\cP
\cQ=-\cQ\cP$ and $\cP q=q\cP$ is the anticommutation of space
reflection operator $\cP$ and sign operator $\cR$:
\be{8}
\cP \cR=-\cR\cP.
\ee
From the fact that $\cR$ and $\cP$ are involutions, $\cR^2=\cP^2=I$,
we find that they can be interpreted as basis (generating)
elements of the real Clifford algebra
\be{8}
R_{2,0}=\span_\RR\{I,\cP,\cR,\cP\cR\}
\ee
or its complex extension
\be{8a}
Cl_2=\span_\CC\{I,\cP,\cR,\cP\cR\}.
\ee
We recall that a real Clifford algebra $R_{m,n}$ with generating
elements $\{e_k\}_{k=1}^{m+n}$
\ba{9}
&&\{e_i, e_k\}:=e_ie_k+e_ke_i=0\quad \forall i\neq k\nn\\
&&e_i^2=I \quad \forall i=1,\ldots, m\nn\\
&&e_i^2=-I \quad\forall i=m+1,\ldots, m+n
\ea
is naturally related to an indefinite form $B(x,y)=\sum_{k=1}^m
x_ky_k-\sum_{k=m+1}^{m+n} x_ky_k$ \ over \ $\RR^{m+n}\ni x,y$
(see, e.g. \cite{cliff-book}, sect. I.1.1). By embedding $R_{m,n}$ into a
complex Clifford algebra, $Cl_{m+n}$,  (complexifying it) the
indefinite metric structure becomes irrelevant and it holds
$R_{m,n}\hookrightarrow R_{m,n}\times \CC\simeq Cl_{m+n}$ for any
metric signature $(m,n)$ with fixed value $m+n$. For $Cl_{m+n}$ it
suffices to work with basis elements of positive type $e_k^2=I,\
\forall k=1,\ldots,m+n$ so that the concrete interpretation as
\rf{8} or \rf{8a} depends only on whether one works with an $\RR-$
or a $\CC-$span.

For a gauged scalar Hamiltonian $H_g$ the Clifford algebra
structures become especially clearly pronounced, e.g., when the
potentials $A(x)$ and $V(x)$ in \rf{2g} under appropriate
regularization are shrunken to an ultra-local support of
delta-function type (see e.g. \cite{kurasov-distrib,AL1,AK}).
Below we demonstrate this fact on a  Hamiltonian with general
regularized zero-range potential at the point $x=0$ as studied,
e.g., in \cite{AL1,AK}
\be{e23}
H_{\mathrm{reg}}=p^2+
t_{11}\la\delta,\cdot\ra\delta+t_{12}\la\delta',\cdot\ra\delta+
 t_{21}\la\delta,\cdot\ra\delta'+t_{22}\la\delta',\cdot\ra\delta'.
\ee
The concrete operator realization ${H}_T$ ($T=\|t_{ij}\|$) in
$L_2(\mathbb{R})$ can be defined by setting
\begin{equation}\label{lesia40}\fl
{H}_T={H}_{\mathrm{reg}}\upharpoonright{\mathcal{D}({H}_T)}, \qquad
 \mathcal{D}({H}_T)=\{\,f\in{W_2^2}(\mathbb{R}\backslash\{0\}) :
 {H}_{\mathrm{reg}}f\in{L_2(\mathbb{R})}\},
\end{equation}
where the derivative $p^2=-\p_x^2$ acts on
${W_2^2}(\mathbb{R}\backslash\{0\})$ in the distributional sense
and the regularized delta-function $\delta$ and its derivative
$\delta'$ (with support at $0$) are defined on the piecewise
continuous functions $f\in{W_2^2}(\mathbb{R}\backslash\{0\})$  as
(for more details see, e.g., \cite{kurasov-distrib})
$$
 \la\delta, f\ra=[f(+0)+f(-0)]/2, \quad \la\delta',
 f\ra=-[f'(+0)+f'(-0)]/2.
$$
Denoting the set of $\cP\cT-$symmetric operators $H_T$,
$[\cP\cT,H_T]=0$, by $\cN_{\cP\cT}$ one immediately verifies that
$H_T\in \cN_{\cP\cT}\quad\Longleftrightarrow\quad
t_{11},t_{22}\in\RR,\quad t_{12},t_{21}\in i\RR$. $\cN_{\cP\cT}$
contains the subset of $\cP-$selfadjoint Hamiltonians which are determined by the condition $t_{12}=t_{21}$. For their $\cP\cT-$symmetric potentials $V=t_{11}\la\delta,\cdot\ra\delta+t_{12}\la\delta',\cdot\ra\delta+
 t_{21}\la\delta,\cdot\ra\delta'+t_{22}\la\delta',\cdot\ra\delta'$ it additionally holds
\be{add1}
\cP V^\dd=V\cP,\qquad \la Vu,v\ra=\la u,V^\dd v\ra,\quad u,v\in
{W_2^2}(\mathbb{R}\backslash\{0\}).
\ee
In analogy to the gauged Hamiltonians \rf{2g}, this
$\cP-$self-adjointness can be modified toward a
$\cP_\phi-$self-adjointness with Clifford-rotated involution
\be{e23a}
\cP_\phi=\cP e^{i\phi\cR}=e^{-i\phi\cR/2}\cP e^{i\phi\cR/2},\qquad
\cR f(x):=\sign(x)f(x)
\ee
so that an appropriate Krein space involution can be constructed
for any parameter combination $t_{12}\neq t_{21}$ as well. The
angle $\phi$ is  fixed by the parameters of the matrix $T$ and can
be defined as follows. One represents $H_{\mathrm{reg}}$ in
\rf{e23} as
\be{e23c}
H_{\mathrm{reg}}f(x)=H_{\mathrm{sym}}^\dd f(x)+(\d(x),\d'(x))(T \G_0
f-\G_1 f),
\ee
where $H_{\mathrm{sym}}^\dd$ is the adjoint of the auxiliary
symmetric operator
\be{e23b-2}\fl
H_{\mathrm{sym}}=-\p_x^2,\qquad
\cD(H_{\mathrm{sym}})=\left\{u(x)\in
{W_2^2}(\mathbb{R}\backslash\{0\})\ | \ u(0)=u'(0)=0\right\}
\ee
and $\G_0$ and $\G_1$ are the average and jump height matrices
\be{e23d}\fl
\G_0f=\frac12\left(
                           \begin{array}{c}
                             f(+0)+f(-0) \\
                             -f'(+0)-f'(-0) \\
                           \end{array}
                         \right),\qquad
\G_1f=\left(
                           \begin{array}{c}
                             f'(+0)-f'(-0) \\
                             f(+0)-f(-0) \\
                           \end{array}
                         \right).
\ee
The domains of $H_T$ in \rf{lesia40} and its adjoint are then
given as \cite{AK}
\ba{e23e}
\cD(H_T)&=&\{f\in {W_2^2}(\mathbb{R}\backslash\{0\})\ | \
T\G_0f=\G_1 f\},\label{e23e-1}\\
\cD(H_T^\dd)&=&\{f\in {W_2^2}(\mathbb{R}\backslash\{0\})\ | \
T^\dd\G_0f=\G_1 f\},\qquad T^\dd:=T^{*T}\label{e23e-2}.
\ea
Taking into account that $\cP_\phi$ commutes with
$H_{\mathrm{sym}}^\dd$ the condition of
$\cP_\phi-$self-adjointness $ \cP_\phi H_T^\dd=H_T\cP_\phi $ is
equivalent to the domain relation $\cP_\phi \cD(H_T^\dd)=\cD(H_T)$
and, hence, also to  $\cP_\phi \cD(H_T)=\cD(H_T^\dd)$. For given
coefficient matrix $T$ these domain relations fix the Clifford
rotation angle $\phi$. Explicitly they imply via \rf{e23e-1} and
\rf{e23e-2}
\ba{e23f}
T\G_0f=\G_1 f,\qquad
T^\dd\G_0\cP_\phi f=\G_1 \cP_\phi f, \qquad f\in\cD(H_T)
\ea
and furthermore via the relations $\left.\cP_\phi f\right|_{x=\pm 0}=\left.e^{\mp i\phi}f\right|_{x=\mp
0}$, $\left.\cP_\phi f'\right|_{x=\pm 0}=\left.-e^{\mp i\phi}f\right|_{x=\mp 0}$ that
\be{e23g}\fl
\begin{array}{rclcrcl}
  \G_0\cP_\phi f&=&M_1\G_0 f + M_2\G_1 f & \qquad & \G_1\cP_\phi f&=&-4M_2\G_0 f + M_1\G_1 f
\end{array}
\ee
where in terms of the Pauli matrices $\sg_1,\sg_2,\sg_3$
\be{e23h}
M_1:=\cos(\phi)\sg_3,\quad M_2:=(i/2)\sin(\phi)\sg_1\,.
\ee
For $f\in \cD(H_T)$ the first relation in \rf{e23f} can be used to
replace $\G_1 f$ by $T\G_0 f$ in \rf{e23g} and to obtain from the
second relation in \rf{e23f} that
\be{e23h}
T^\dd M_2 T=M_1T-T^\dd M_1-4M_2.
\ee
Due to the $\cP\cT-$symmetry-induced structure of $T$, i.e.
$t_{11},t_{22}\in \RR$, $t_{12},t_{21}\in i\RR$ this matrix
equation strongly simplifies and yields the defining
relation for the Clifford-rotation angle $\phi$
\be{e23i}
i\sin(\phi)\left[\det (T)+4\right]=2\cos(\phi)(t_{12}-t_{21}).
\ee
For this specific angle $\phi$ the $\cP\cT-$symmetric Hamiltonian
$H_T$  in \rf{lesia40} is $\cP_\phi-$self-adjoint, $\cP_\phi
H_T^\dd=H_T\cP_\phi$. Accordingly, for the $\cP\cT-$symmetric potential $V$ it holds (conf. \rf{add1})
\be{e24}
\cP_\phi V^\dd=V\cP_\phi, \qquad \la V u,v\ra=\la u,V^\dd v\ra,\quad u,v\in
{W_2^2}(\mathbb{R}\backslash\{0\})
\ee
with the rotated involution $\cP_\phi=e^{-i\phi\cR/2}\cP e^{i\phi\cR/2}$ built
from the Clifford algebra  elements (involutions) $\cP$ and $\cR$. In the special case of $\phi=0$ eq. \rf{e23i} implies $t_{12}=t_{21}$ so that \rf{e24} indeed coincides with \rf{add1}, and $\cP_{\phi=0}=\cP$.\\[1ex]

\noindent
{\bf Concluding remarks}
 \begin{itemize}
   \item The Cartan decomposition used here for the
   structure analysis of the gauge potentials $A$ can also be
   applied to the similarity
   transformation\footnote{We use the notations from \cite{B4,B-C,AGK,M-metric}
   with $\rho^2=e^{-Q}=\cP\cC$ and the $\cC-$operator, as usual,
   as dynamical symmetry $[\cC,H]=0$ and involution $\cC^2=I$.} $\rho$
   which maps  a spectrally diagonalizable $\cP\cT-$symmetric Hamiltonian $H$ with real spectrum
   into its equivalent Hermitian operator $h=\rho H\rho^{-1}$.  Although,
   in general, $\rho$ is a highly nonlocal operator, as similarity
   transformation it can nevertheless be understood as Lie group element.
   Within the framework of generalized Cartan decompositions the
   Hermiticity $\rho=\rho^\dd$ and positivity $\rho>0$ clearly
   indicate that $\rho$ should be an element of some noncompact
   coset space. For the simple finite-dimensional matrix setups
   of \cite{GS-pra,GS-prl,gain-loss1} this non-compactness
   of $\rho$ was clearly visible in its $SO(m,\CC)$ "boost"-type.
   \item The possible use of the generalized
   Jaynes-Cummings setup of \cite{transmon1,transmon2}
   as reliable experimental candidate for the implementation of qubit states,
   together with the structural links indicated here, seems to open a new
   and interesting playground for experimental implementations of $\cP\cT-$symmetric
   and Lie-triple setups as well.
   \item The symmetric operator $H_{\mathrm{sym}}$ in \rf{e23b-2} commutes with both generating involutions $\cP$ and $\cR$ from the Clifford algebra $Cl_2$ in \rf{8a}. It will be shown in \cite{KT}  that for any involution $J$ constructed in an arbitrary way from $Cl_2-$involution elements there necessarily exists a very special subclass of $J$-self-adjoint extensions of $H_{\mathrm{sym}}$ which will have a spectrum filling the whole complex plane $\CC$.
   \item It is known (see, e.g., sect. I.3.5 in \cite{cliff-book}) that
   a Clifford algebra $Cl_m$ with $m$ basis elements $\left\{e_1,\cdots,e_m\right\}$
   has a faithful representation as matrix algebra $Cl_{2k}\sim \CC^{2^k\times 2^k}$, \ $ Cl_{2k+1}\sim \CC^{2^k\times
2^k}\oplus \CC^{2^k\times 2^k} $. Furthermore, it is known that
the $J-$selfadjoint extensions of a symmetric operator with
deficiency indices $\la n, n\ra$ are parameterized by unitary
matrices $U\in U(n)\subset \CC^{n\times n}$. Once, the
extension-related Clifford elements act via a representation in
this $\CC^{n\times n}$ matrix space the maximal number $m$ of
Clifford basis elements in $Cl_m$ is bounded by the dimensionality
of this matrix space and, hence, by $2^k\le n$ for $m=2k$ and
$2^{k+1}\le n$ for $m=2k+1$. The Hamiltonian $H_T$ in \rf{lesia40}
is related to the symmetric operator $H_{\mathrm{sym}}$
in \rf{e23b-2} with deficiency indices $\la 2, 2\ra$ and parameter
matrix $U\in U(2)$ \cite{AGK}. This means that not more than the
two Clifford basis elements $\cP$ and $\cR$ can be naturally
associated with this operator extension.
 \end{itemize}

\ack
UG thanks Steven Duplij for useful discussions on $n-$ary Lie algebras and DFG for support within the Collaborative Research Center SFB 609. SK acknowledges support by DFFD of Ukraine (F28.1/017) and
JRP IZ73Z0 of SCOPES 2009-2012.

 \end{document}